\documentclass{emulateapj}
\usepackage{amssymb,amsmath}
\usepackage{color,hyperref}
\definecolor{linkcolor}{rgb}{0,0,0.25}
\hypersetup{
  colorlinks=true,        % false: boxed links; true: colored links
  linkcolor=linkcolor,    % color of internal links
  citecolor=linkcolor,    % color of links to bibliography
  filecolor=linkcolor,    % color of file links
  urlcolor=linkcolor      % color of external links
}
\newcounter{address}
\newcommand{\ie}{i.e.}
\newcommand{\etal}{et al.}
\newcommand{\dd}{\mathrm{d}}
\newcommand{\eg}{e.g.}
\newcommand{\eqnname}{equation}

\newcommand{\sectionname}{$\mathsection$}
\newcommand{\segue}{\emph{SEGUE}}
\newcommand{\sdss}{\emph{SDSS}}

\newcommand{\flag}[1]{\texttt{\lowercase{#1}}}

\newcommand{\feh}{\ensuremath{[\mathrm{Fe/H}]}}
\newcommand{\afe}{\ensuremath{[\alpha\mathrm{/Fe}]}}

\newcommand{\age}{\ensuremath{\tau}}

\newcommand{\bovyetal}{B11}

\begin{document}

\submitted{Astrophys.~J., in press}

\title{The Milky Way has no thick disk}
\author{Jo~Bovy\altaffilmark{1,2,3},
  Hans-Walter~Rix\altaffilmark{4}, 
  \& David~W.~Hogg\altaffilmark{4,5}}
\altaffiltext{\theaddress}{\stepcounter{address} Institute for Advanced Study, Einstein Drive, Princeton, NJ 08540, USA}
\altaffiltext{\theaddress}{\stepcounter{address} Hubble fellow}
\altaffiltext{\theaddress}{\stepcounter{address}
  Correspondence should be addressed to bovy@ias.edu~.}
\altaffiltext{\theaddress}{\stepcounter{address}
  Max-Planck-Institut f\"ur Astronomie, K\"onigstuhl 17, D-69117
  Heidelberg, Germany}
\altaffiltext{\theaddress}{\stepcounter{address} Center for
  Cosmology and Particle Physics, Department of Physics, New York
  University, 4 Washington Place, New York, NY 10003, USA}

\begin{abstract} 
  Different stellar sub-populations of the Milky Way's stellar disk
  are known to have different vertical scale heights, their thickness
  increasing with age. Using \segue\ spectroscopic survey data, we
  have recently shown that mono-abundance sub-populations, defined in
  the \afe-\feh\ space, are well described by single exponential
  spatial-density profiles in both the radial and the vertical
  direction; therefore any star of a given abundance is clearly
  associated with a sub-population of scale height $h_z$. Here, we
  work out how to determine the stellar surface-mass density
  contributions at the solar radius $R_0$ of each such sub-population,
  accounting for the survey selection function, and for the fraction
  of the stellar population mass that is reflected in the
  spectroscopic target stars given populations of different abundances
  and their presumed age distributions.  Taken together, this enables
  us to derive $\Sigma_{R_0}(h_z)$, the surface-mass contributions of
  stellar populations with scale height $h_z$.  Surprisingly, we find
  no hint of a thin-thick disk bi-modality in this mass-weighted
  scale-height distribution, but a smoothly decreasing function,
  approximately $\Sigma_{R_0}(h_z)\propto \exp(-h_z)$, from $h_z
  \approx 200$ pc to $h_z \approx 1$ kpc. As $h_z$ is ultimately the
  structurally defining property of a thin or thick disk, this shows
  clearly that the Milky Way has a continuous and monotonic
  distribution of disk thicknesses: there is no `thick disk' sensibly
  characterized as a distinct component.  We discuss how our result is
  consistent with evidence for seeming bi-modality in purely geometric
  disk decompositions, or chemical abundances analyses. We constrain
  the total visible stellar surface-mass density at the Solar radius
  to be $\Sigma^{^*}_{R_0} = 30 \pm 1\ M_\odot$ pc$^{-2}$.
\end{abstract}


\section{Introduction}

In both the Milky Way \citep{Gilmore83a} and external galaxies
\citep{Tsikoudi79a,Burstein79a}, the stellar-density or luminosity
profiles perpendicular to the disk plane reveal an excess over a
simple exponential ``thin disk'' at distances $|z| \gtrsim$ 1 kpc.
This excess, confirmed by many studies in the last two decades
\citep[\eg,][]{Reid93a,Juric08a}, seems to be a generic feature of
galactic disks \citep[\eg,][]{Yoachim06a} and has typically been
described by, and often ascribed to, the presence of a ``thick-disk''
component, distinct from both the halo and thin disk components
\citep[\eg,][]{Juric08a}. However, whether or not the excess is
sensibly attributed to a distinct ``thick-disk'' component is, in our
view, an open question (see \sectionname~\ref{sec:outline}).

Studies of the age, kinematics, and elemental-abundance ratios of
probable members of the thicker disk component have revealed that it
is older \citep[\eg,][]{Bensby05a}, kinematically hotter
\citep[\eg,][]{Chiba00a}, and metal-poor and enhanced in $\alpha$
elements \citep[\eg,][]{Fuhrmann98a,Prochaska00a,Lee11a}. Recent
analyses \citep{Navarro11a,Lee11a} also revealed a striking bi-modal
distribution of disk stars in the elemental-abundance parameter plane
of \afe\ and \feh.  However, these analyses did not account for proper
volume sampling, and the \afe\ distribution will appear bi-modal even
if the underlying (enrichment) age distribution is perfectly smooth
\citep{Schoenrich08a}, simply because \afe\ strongly changes as soon
as enrichment by Type Ia supernovae (SNe Ia) becomes
important. 

These conceptual issues can be illustrated by the recent finding that
geometric decompositions of the Galactic disk yield strikingly
different results for the structural parameters (\eg, the radial scale
lengths) for the thicker and thinner components, when they are based
on an elemental-abundance selection of sample stars
\citep[][\bovyetal\ hereafter]{Bensby11a,Bovy11a}, as opposed to
luminosity- or volume-selection \citep[\eg,][]{Juric08a}.  This is
because sample selection by elemental abundance, \eg, \afe\ and \feh,
can link sub-sample members independent of their position and
velocity, while selections based only on luminosity, volume or
kinematics correlate the sample selection with the structure to be
inferred.  Therefore, discerning whether the thin- and thick-disk
components should be thought of as distinct has been difficult on the
basis of the existing evidence, given the modest number statistics
and, most importantly, the conceptual or practical selection effects
of many observational studies. Such issues have troubled the question
of a thin--thick disk dichotomy for a long time
\citep[\eg,][]{Nemec91a,Nemec93a,Ryan93a,Norris99a}.

The question of a thick-disk component plays an important part in our
understanding of how our Galaxy formed and evolved. Various
qualitatively dissimilar scenarios have been proposed to explain the
presence of a thick-disk component. Many of these invoke mechanisms
external to the already existing disk that may be expected in
hierarchical structure formation: direct accretion of stars from a
disrupted satellite galaxy \citep{Abadi03a}, heating of a pre-existing
thin disk through a minor merger
\citep[\eg,][]{Quinn93a,Kazantzidis08a}, or star formation induced by
a gas-rich merger \citep[\eg,][]{Brook04a}. However, quiescent
internal dynamical evolution can also reproduce the locally observed
properties of the Milky Way's thick disk component
\citep[\eg,][]{Schoenrich08a,Schoenrich09a}. Recent attempts to
differentiate these scenarios through volume-averaged stellar
kinematics have not been conclusive
\citep[\eg,][]{Sales09a,Dierickx10a,Wilson11a}.

Discerning the mechanism that leads to the thickness distribution of
galactic disks is important in constraining the rate of minor mergers,
the resilience of stellar disks to such mergers, and the importance of
`internal' heating mechanisms such as radial migration
\citep{Sellwood02a} or turbulent, gravitationally unstable disk
evolution \citep[\eg,][]{Bournaud09a,ForsterSchreiber09a}. While external
heating mechanisms or distinct disk-formation epochs may lead to
sensibly distinct `thin' and `thick' disk components, internal
mechanisms should not. Ultimately, these questions are also important
for understanding how much of an archaeological record our Galactic
disk holds and how much formation memory has been erased
\citep{Freeman02a}.

\subsection{Approaches to dissecting the Galactic
  disk}\label{sec:outline}

In our view there exists no prior analysis that can answer the
question of whether it is sensible to view the thinner and thicker
parts of the Milky Way's disk as distinct components rather than a
smooth distribution of stellar-disk scale heights. This is mainly for
three reasons:

$\bullet$ Either analyses have started out with insisting that two
vertically-exponential components be fit to the observations; this may
be a sensible approach if the data are not good enough to warrant more
complex models, but it pre-supposes the answer. \bovyetal\ showed that
chemically defined sub-populations correspond to simple,
single-exponential, sub-components of the Milky Way's disk. \bovyetal\
also found a wide range of vertical scale height for these
chemically-defined sub-populations, smoothly varying from thinner to
thicker components with increasing (enrichment) age. 

$\bullet$ Decompositions have been based on geometric or kinematic
sample definitions. Absent a very clear separation of components in
position or velocity space, this approach inevitably has an air of
circular reasoning that precludes unique decompositions. Indeed, our
recent analysis (\bovyetal) has shown that state-of-the-art geometric
decompositions \citep[\eg,][]{Juric08a} cannot even reliably tell
whether the thicker disk components are radially more extended or more
concentrated, as a comparison with a structure-independent
abundance-based sub-sample selection has shown. Studies of the
elemental-abundance trends of kinematically-selected samples of stars
\citep[\eg,][]{Prochaska00a,Feltzing03a} cannot assess the
distinctness of the thick-disk component without correcting for the
(strong) biases induced by the selection. The kinematics and spatial
structure of a population of stars are inextricably coupled through
the dynamical properties of the population, and are therefore not
\emph{a priori} independent.

\begin{figure*}[t]
\includegraphics[width=0.5\textwidth,clip=]{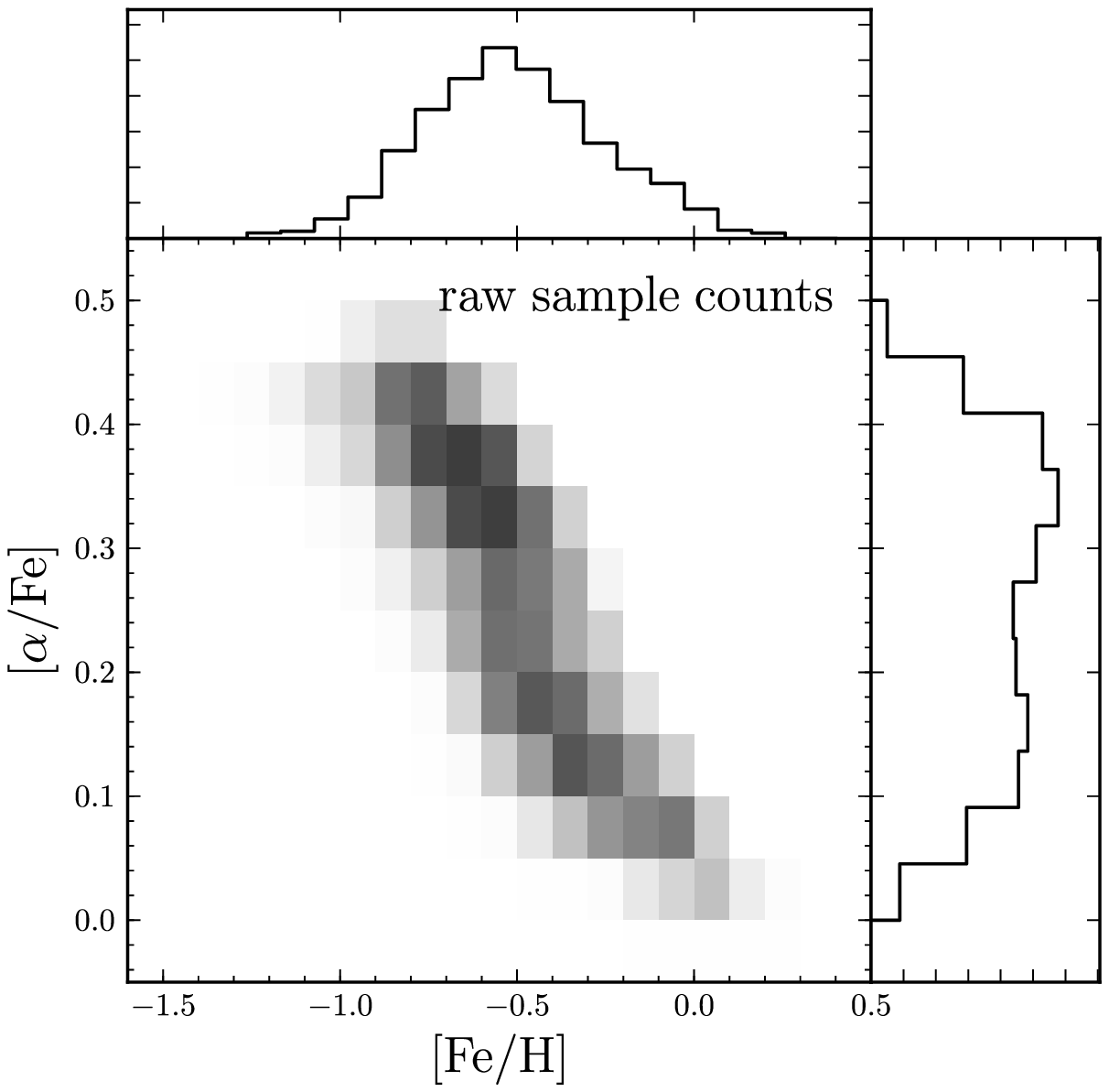}
\includegraphics[width=0.5\textwidth,clip=]{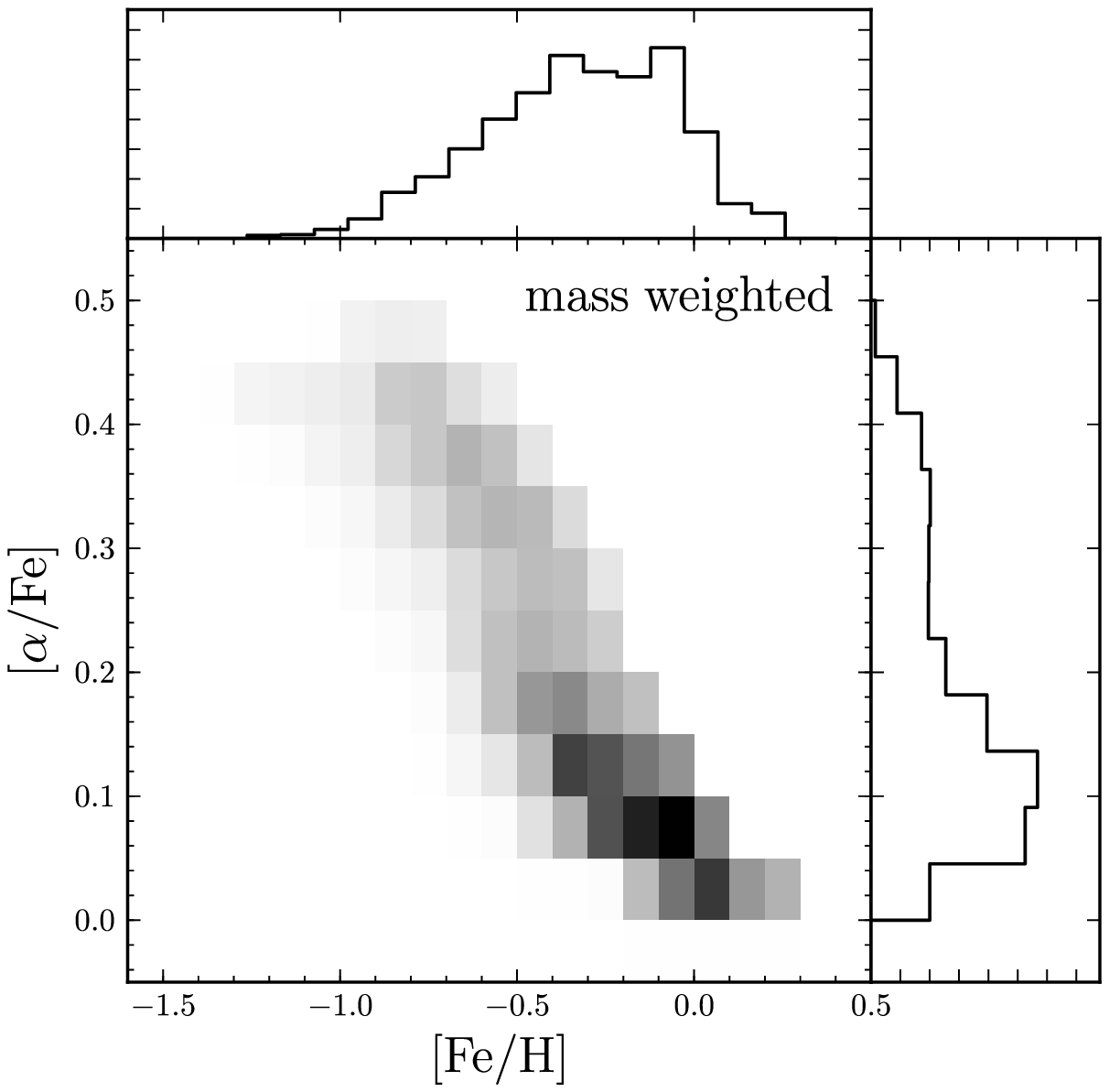}
\caption{Distribution of G-type dwarfs in the
  \sdss/\segue\ spectroscopic sample in the
  \feh,\afe\ elemental-abundance plane. The left panel shows the raw
  number counts, while the right panel shows mass-weighted number
  counts that are corrected for spectroscopic selection effects and
  converted into total stellar surface-mass densities at the Solar
  radius using the stellar population modeling described in
  \appendixname~\ref{sec:surfmass}. Pixels span 0.1 dex in \feh\ and
  0.05 dex in \afe, which is larger than in the equivalent Figure in
  \citet{Lee11a}, as the spatial number-density fitting used as part
  of the mass-weighting demands larger bins. The mass weighting shows
  that the prominent bi-modality seen in the raw number counts
  \citep[left, cf.][]{Lee11a} is mostly a consequence of the uneven
  spectroscopic sampling of the underlying stellar populations. The
  remaining hint of \afe\ bi-modality seen in the right panel is a
  natural consequence of the enrichment physics, \ie, SN Ia enrichment
  delays, even for a perfectly uniform (enrichment) age distribution
  \citep[\eg,][]{Schoenrich08a}.}\label{fig:mass_afe_feh}
\end{figure*}

$\bullet$ Conceptually the right approach seems to be to define
sub-samples by a property that may correlate with disk structure but
is formally independent of it, such as stellar age or stellar
abundances. Such `tags' are properties that do not change along the
orbit of the star, nor do they change if a star changes its orbit due
to a minor merger or radial migration \citep{Freeman02a}; in this
sense, tagging stellar populations by elemental abundances is even
better than using the best structural or dynamical tags available:
integrals of motion, or orbital actions. With such sub-samples, one
can then ask whether their respective disk structures suggest distinct
thin and thick disks.  This approach has been re-advocated in recent
studies \citep{Navarro11a,Lee11a}, drawing on the seemingly bi-modal
distribution of stars in the \afe\ - \feh\ plane of elemental
abundances; however, these studies did not account for selection
effects and volume corrections.

Believing that the last approach is the conceptually correct one, we
derive in this paper for the first time a scale-height distribution of
the stellar mass at the Solar radius in the Galaxy to see whether this
distribution shows any evidence of bi-modality. This requires two
steps: first, one needs to associate a vertical scale-height with any
given star. Our recent analysis (\bovyetal) showed that
sub-populations of stars of a given \afe\ and \feh\ are well described
by a single exponential in their vertical profile, making (\afe ,\feh
) a suitable tag for each star that uniquely determines the
scale-height of its sub-population. Second, one needs to determine
what the (surface) mass fraction of stars is in this \afe --
\feh\ sub-population.  As such samples can only be defined through
spectroscopic surveys, this in turn requires proper accounting for the
spectroscopic survey selection function (\bovyetal), and an estimate
of the stellar mass fraction that the spectroscopically targeted stars
represent of the entire population at this elemental abundance.

We work out this last step in this paper
(\sectionname~\ref{sec:main}), and on that basis derive the
distribution of stellar surface-mass density-weighted distribution of
stellar scale heights at $R_0$, $\Sigma_{R_0}(h_z)$
(\sectionname~\ref{sec:main}). Surprisingly, we find a smooth,
exponentially-declining surface-mass-density spectrum as a function of
scale height between $\approx$ 200 and 1000 pc. This mass spectrum
shows no gaps, excess, or hints of bi-modality at large scale heights,
leading us to conclude that the Milky Way has no distinct thick
disk. We discuss implications, and explain how this relates to recent
work that arrived at qualitatively different conclusions, in
\sectionname~\ref{sec:discuss}.

\section{The mass-weighted scale height distribution of disk stars in the Milky Way}\label{sec:main}

Recently, we have performed (in \bovyetal) number-density fits to
sub-populations of stars defined as narrow boxes in the
elemental-abundance plane spanned by metallicity \feh\ and
$\alpha$-enhancement \afe, based on G-dwarf spectra and star counts
from \sdss/\segue\ (\citealt{Abazajian09a,Yanny09a}; \segue\
elemental-abundance uncertainties are 0.2 dex in \feh\ and 0.1 dex in
\afe). G-type dwarfs are the most luminous tracers whose main-sequence
lifetime is larger than the expected disk age at basically all
metallicities. We refer the reader to \bovyetal\ for a discussion of
the data set and for a detailed description and discussion of these
fits, accounting for the volume correction of the spectroscopic
survey. \bovyetal\ show that each mono-abundance sub-population has a
simple density structure that can be described by an exponential
profile both in Galactocentric radius $R$ and vertical height $z$. The
inferred scale heights for different mono-abundance sub-populations
vary with \afe\ and \feh , increasing smoothly from about 200 pc to
about 1200 pc when going from populations with near-solar abundances
to populations that are more metal-poor and enhanced in $\alpha$
elements.

These results were derived from the spectroscopic \segue\ G-dwarf
sample, whose face-value distribution in the \feh--\afe\ plane is
suggestive of two distinct populations definable in this
elemental-abundance space, because there is a peak near solar
abundances and one at metal-poor and $\alpha$-enhanced abundances (see
\citealt{Lee11a}, \bovyetal, and the left panel of
\figurename~\ref{fig:mass_afe_feh}).  However, \segue\ targeted stars
at high latitudes in an apparent-magnitude range $14.5 \leq r \leq
20.2$ that finds G-type dwarfs at vertical heights $\gtrsim 500$ pc,
significantly above the bulk of the thin-disk population (which
\bovyetal\ inferred to have a scale height of 250 pc). Additionally,
G-type dwarfs of different metallicities have different luminosity and
hence survey volumes, and the spectroscopic targeting was weighted
toward the fainter end of the apparent-magnitude range and toward
higher latitudes (see \citealt{Yanny09a} and the discussion in
\appendixname~A of \bovyetal). Thus, the distribution of the \segue\
sample in the \feh--\afe\ plane reflects the survey selection function
more than the abundance distribution of the Milky Way stars within 1
to 2 kpc from the Sun.

To properly assess the contribution of the various mono-abundance
sub-populations to the mass or surface-mass budget in the Solar
neighborhood, we must convert the number of spectroscopically-observed
stars in each bin to a surface-mass density at the Solar neighborhood.
On the one hand, this requires incorporation of a model for the
\segue\ selection function (see \appendixname~A of \bovyetal) and our
exponential-disk fits (\bovyetal). On the other hand, this requires
the use of stellar-population models that can relate the observed
number of G-type dwarfs to the total mass of the stellar population,
given the metallicity of the sub-population and an assumed star
formation history for it. This is described in detail in
\appendixname~\ref{sec:surfmass}.

Briefly, for G-type dwarfs in a given (\feh,\afe) bin we calculate
their total number per square pc (integrated over vertical height) by
adjusting the normalization of the number-density profile such that
after running it through our model for the \segue\ selection function
it predicts the observed number of stars in each bin. Then we relate
the number density of G-type dwarfs to the total stellar surface-mass
density by multiplying the number density by the average mass of a
G-type dwarf---calculated using Padova isochrones
\citep{Marigo08a}---and dividing it by the fraction of the mass in a
stellar population in G-type dwarfs (calculated using the same
isochrones and assuming a lognormal \citet{Chabrier01a} initial mass
function). At a given abundance, this fraction of course depends on
the age of the population, and here is calculated by marginalizing
over a flat age distribution between 0.5 and 10 Gyr for each
bin. However, averaging only over older ages for $\alpha$-enhanced
stars would be appropriate, as $\alpha$-enhanced stars likely
represent the oldest part of the disk. As we show in
\appendixname~\ref{sec:surfmass}, this gives similar results, with a
slightly steeper decline in $\Sigma(R_0)$ with $h_z$. We calculate
uncertainties on the surface-mass densities by varying the
density-parameters according to the posterior probability distribution
for the parameters in \bovyetal. These uncertainties do not include
systematic uncertainties due to the use of the stellar isochrones.
This procedure results in an estimate of the stellar surface-mass
density contribution at the Solar radius for any abundance-selected
sub-population, which in turn has a vertical scale height associated
with it. The relative total stellar surface-mass densities of
different mono-abundance bins are not affected by assuming a different
initial mass function (IMF), although assuming a different IMF can
systematically shift all surface-mass densities by a few percent (see
below).

\begin{figure}[t]
\includegraphics[width=0.5\textwidth,clip=]{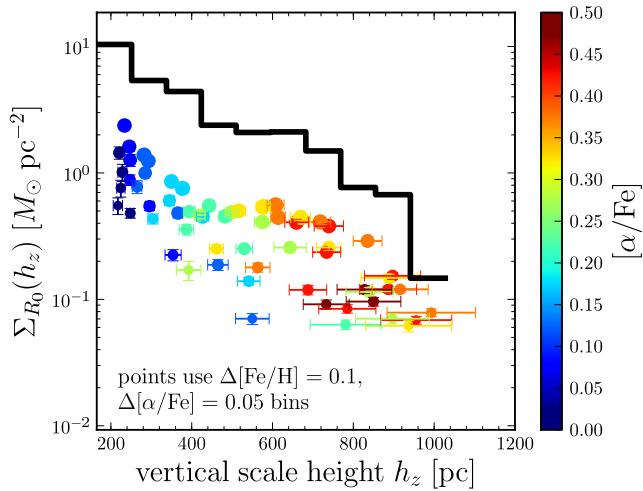}
\caption{Distribution of stellar surface-mass density at the Solar
  radius $\Sigma_{R_0}(h_z)$ as a function of vertical scale height
  $h_z$.  The thick black histogram shows the total stellar
  surface-mass density in bins in $h_z$, calculated by summing the
  total stellar masses of sub-populations in bins in \afe\ and \feh.
  The stellar surface-mass densities of the individual
  elemental-abundance bins in \feh\ and \afe\ are shown as dots, with
  values for $\Sigma_{R_0}(\feh,\afe)$ on the $y$-axis. The points are
  color-coded by the value of \afe\ in each bin and the size of the
  points is proportional to the square root of the number of data
  points that the density fits are based on. Some of the errorbars are
  smaller than the points. Elemental abundance bins have a width of
  0.1 in \feh\ and 0.05 in \afe.}\label{fig:mass_hz_afe}
\end{figure}

The results from this mass estimation are shown in \figurename
s~\ref{fig:mass_afe_feh} and \ref{fig:mass_hz_afe}.  The left panel of
\figurename~\ref{fig:mass_afe_feh} is simply a more coarsely binned
version of the unweighted \segue\ G-dwarf sample abundance
distribution, which shows two distinct maxima, one considerably more
metal poor and $\alpha$-enhanced than the other, seemingly reflecting
a chemically-distinct thick-disk component. It is important to note
that the marginalized \feh\ metallicity distribution (left panel, top)
shows no hint of any bi-modality. There is distinct bi-modality in the
marginalized \afe\ distribution, but \citet{Schoenrich08a} already
showed that even for a smooth age distribution such bi-modality
arises, simply separating stars that formed before and after
enrichment by SN Ia became important. The right panel of
\figurename~\ref{fig:mass_afe_feh} shows the stellar surface-mass
density at the solar radius $R_0$ in each elemental-abundance bin,
corrected for selection effects due to the spectroscopic
\segue\ selection as described above.  It represents the properly
mass-weighted, underlying distribution of disk stars in the
elemental-abundance space spanned by \feh\ and \afe, and dramatically
differs from the raw sample distribution in the left panel: it does
not have the strong bi-modality apparent in the raw
\sdss/\segue\ number distribution. It also shows no hint of a bi-modal
\feh\ metallicity distribution and the remaining hint of
\afe\ bi-modality is explained as in the left panel.

The right panel of \figurename~\ref{fig:mass_afe_feh} now provides the
relevant weights of each mono-abundance bin to a surface-mass-weighted
distribution of disk scale heights.  The colored symbols in
\figurename~\ref{fig:mass_hz_afe} show exactly these surface-mass
density contributions for all these mono-abundance bins in \feh\ and
\afe\ (with widths 0.1 in \feh\ and 0.05 in \afe) versus the scale
height of those sub-population from \bovyetal, color-coded by their
\afe\ enhancement. We then sum the surface-mass contributions
$\Sigma_{R_0}(h_z|\feh,\afe)$ of sub-populations (\feh,\afe) into bins
in scale height. This results in the thick black histogram, which
represents $\Sigma_{R_0}(h_z)$, or more simply $p(h_z)$, the
surface-mass weighted distribution of vertical scale heights in the
Solar neighborhood (which we will refer to in the remainder simply as
the `scale-height distribution'). That is, for any random stellar-mass
element this function gives the probability density for the scale
height of the structural component to which it belongs. This is the
function we set out to construct in order to examine whether it makes
sense to think of distinct thin and thick disk components in the Milky
Way. Remarkably, we find that the scale-height distribution simply
decreases quite smoothly towards larger scale heights, with an
approximately exponential relation between surface-mass density and
scale height $\Sigma(R_0) \propto \exp(-h_z)$. The scale height
distribution does not show any gaps, excesses, or hints of
bi-modality, beyond this simple relation.

By combining all of the stellar surface-mass density estimates we can
precisely measure the total visible stellar surface-mass density at
the Solar radius. We find $\Sigma^{^*}_{R_0} = 30 \pm 1\ M_\odot$
pc$^{-2}$. This is similar to the estimate of \citet{Flynn06a}, who
report $\Sigma^{^*}_{R_0} = 29\ M_\odot$ pc$^{-2}$. This estimate
depends slightly on the assummed IMF. Using the exponential IMF (IMF3)
of \citet{Chabrier01a} gives $\Sigma^{^*}_{R_0} = 29.5\ M_\odot$
pc$^{-2}$; the IMF from \citet{Chabrier03a} gives $\Sigma^{^*}_{R_0} =
29\ M_\odot$ pc$^{-2}$; and a \citet{Kroupa03a} IMF gives
$\Sigma^{^*}_{R_0} = 32\ M_\odot$ pc$^{-2}$.

\section{Discussion}\label{sec:discuss}

\figurename~\ref{fig:mass_afe_feh} shows that properly correcting for
the spectroscopic sampling of the underlying stellar sub-populations
is crucial in assessing the elemental-abundance distribution at $R_0$
of spectroscopically selected samples of stars. The abundance
distribution without this correction is heavily influenced by the
survey-specific spatial and mass sampling of the underlying stellar
population---both of which act to make the metal-poor and
$\alpha$-enhanced sub-populations more prominent in the high-latitude
and color-selected \segue\ sample---which leads to a spuriously
enhanced bi-modality in elemental-abundance space. The mass-weighted
metallicity distribution in \figurename~\ref{fig:mass_afe_feh} has no
bi-modality. The mass-weighted \afe-distribution in the same Figure
has only a hint of a bi-modality, as is expected in standard smooth
star formation and enrichment scenarios \citep{Schoenrich08a}, and
reflects enrichment physics and not galaxy evolution.

Improper weighting of the distribution of structural parameters can
again lead to spurious bi-modal signatures. \figurename~5 in
\bovyetal\ shows the location of chemically-defined sub-populations in
the space of the structural parameters (radial scale length, vertical
scale height). This scatterplot, based on equal-area bins in
(\feh,\afe), gives undue prominence to the low-\feh, high-\afe\ (with
respect to solar abundances) bins, as many of them only contribute a
negligible amount to the total stellar mass. Here,
\figurename~\ref{fig:mass_hz_afe} shows that the proper mass-weighting
of the (\feh,\afe) sub-populations gives a vertical-scale-height
distribution that is smooth and monotonically declining between
thinner disk component scale heights of 200 pc and thicker components'
scale heights of $\approx 1200$ pc, with no gaps or excesses beyond
the smooth, approximately exponential distribution. \bovyetal\ found
that each elemental-abundance bin was preferentially fit by a single
exponential rather than by two disk components, such that the smooth
scale-height distribution in \figurename~\ref{fig:mass_hz_afe} is
\emph{not} merely the result of the smoothing out of an intrinsically
bi-modal distribution by elemental-abundance errors (which are small
for the \segue\ sample, see \sectionname~\ref{sec:main}) or
overlapping abundance distributions of distinct ``thin'' and ``thick''
disks; if either of these were the case B11 should have detected two
components in the abundance bins with single-exponential scale heights
in the range of approximately 400 to 600 pc. The large uncertainties
on the radial scale lengths of the chemically-defined mono-abundance
sub-populations in \bovyetal\ complicate a similar assessment of the
radial structure of the disk, but this has no bearing on the analysis
of the vertical structure. Upcoming surveys such as \emph{APOGEE}
\citep{Eisenstein12a} or \emph{HERMES/GALAH} \citep{Freeman10a} that
will sample stellar populations in the plane of the Milky Way will be
able to study the radial structure in more detail.

Thus, stars in the Solar neighborhood have a smoothly decreasing
probability of belonging to structural components with increasing
scale heights. This implies that the thicker disk component in the
Milky Way is simply the tail of a continuous and monotonic
scale-height distribution. This has been suggested before
\citep[\eg,][]{Norris87a,Schoenrich09a}, but never directly measured
as we do here. As such, there is no distinct thick-disk component in
our Galaxy. 

Together with the findings in \bovyetal\ that the thicker and older
components of the Galactic disk have a shorter radial scale lengths
than the thinner and younger components this qualitatively points
toward a continuous internal mechanism such as radial migration or
turbulent disk evolution being predominantly responsible for the
thickening of the disk, rather than an external merger or heating
event. However, a rigorous comparison with models, where thick
stellar-disk components arise from one or a few distinct events
triggered, \eg, by satellite infall, is needed to see whether the
present data are indeed inconsistent with the results presented
here. Formation age of stars may serve as a sensible marker in
simulations to associate individual stars with a `parent
sub-population' whose scale height can be determined. In this context,
it is worth noting that combined data from \emph{APOGEE} and
\emph{Kepler} will soon provide ages for a large number of stars
through asteroseismology \citep[\eg,][]{Gilliland10a}, which could
help in mapping mono-age populations into the mono-abundance
populations studied here.

The proper selection function analysis and subsequent conversion to
stellar mass densities can of course be used in the future to look
with proper mass weighting at other disk diagnostics beyond the
spatial distribution, such as the orbital eccentricity or vertical
motions, which has not been done correctly
\citep[\eg,][]{Dierickx10a,Wilson11a}.

Our results that show that the Milky Way has no distinct thick-disk
component might appear to be at odds with observations of external
edge-on galaxies, where ``thick disk'' components are found to be
universal \citep[\eg,][]{Yoachim06a}. However, those decompositions
are restricted to luminosity-weighted geometric decompositions and
suffer from the uncertain influence of dust in the mid-plane of the
galaxies; they cannot isolate components based on elemental
abundances; and they only perform discrete two-component fits, which
naturally prefers two components over a single exponential component,
even though the underlying scale-height distribution might be more
complicated, as shown in this paper. The findings from external
galaxies might be more correctly interpreted as proving the
universality of \emph{thicker disk components} rather than that of
``thick disks''.

\acknowledgements It is a pleasure to thank Tim Beers, Eric Bell, Dan
Foreman-Mackey, Ken Freeman, Gerry Gilmore, Patrick Hall, George Lake,
Young Sun Lee, Chao Liu, Steven Majewski, Julio Navarro, Connie
Rockosi, Scott Tremaine, Glenn van de Ven, and Lan Zhang for helpful
comments and assistance. Support for Program number HST-HF-51285.01
was provided by NASA through a Hubble Fellowship grant from the Space
Telescope Science Institute, which is operated by the Association of
Universities for Research in Astronomy, Incorporated, under NASA
contract NAS5-26555.  J.B. and D.W.H. were partially supported by NASA
(grant NNX08AJ48G) and the NSF (grant AST-0908357).  D.W.H. is a
research fellow of the Alexander von Humboldt Foundation of Germany.
J.B. and H.W.R. acknowledge partial support from SFB 881 funded by the
German Research Foundation DFG.

Funding for the SDSS and SDSS-II has been provided
by the Alfred P. Sloan Foundation, the Participating Institutions, the
National Science Foundation, the U.S. Department of Energy, the
National Aeronautics and Space Administration, the Japanese
Monbukagakusho, the Max Planck Society, and the Higher Education
Funding Council for England. The SDSS Web Site is
http://www.sdss.org/.

\appendix

\section{The surface-mass density in a mono-abundance sub-population}\label{sec:surfmass}

In practice, we calculate the surface-mass density contribution at the
Solar radius from each mono-abundance sub-component, defined as a box
in the \feh--\afe\ plane, by first converting the observed number
counts into a `number column density' ($N(R_0)$ in the notation of
\bovyetal; units \flag{stars pc$^{-2}$}) and then converting this
number density into a total stellar surface-mass density using
stellar-population models. This involves estimating the mean
(individual) stellar masses of the stars we have in the sample, and
estimating which fraction of their total stellar population at that
abundance that mass range constitutes.

When performing the stellar number-density fits in \bovyetal\ we did
not fit for the normalization of the density. However, we can
calculate this normalization by adjusting it such that it predicts the
observed number of stars when we run the density model through the
observational selection function.  Thus, we calculate the predicted
number of stars for a density normalized to have unit surface-mass
density (in \flag{stars pc$^{-2}$}); the correct normalization
constant is then the total number of observed stars divided by the
predicted number of stars for unit surface-mass density. We calculate
the uncertainty in this number by performing this procedure for each
of the samples from the posterior distribution for the number-density
parameters, obtained in \bovyetal\ to estimate the uncertainties in
the best-fit number-density parameters. This gives for each abundance
bin a number $N(R_0;\feh,\afe) \pm \sigma_{N;\feh,\afe}$ that
represents the total number of stars per square pc in this \feh--\afe\
bin at the Solar radius.

In order to turn this number density $N(R_0;\feh,\afe)$ into a stellar
mass we use stellar isochrones in the SDSS photometric system
\citep{Girardi04a,Marigo08a,Girardi10a}\footnote{Retrieved using the
  Web interface provided by Leo Girardi at the Astronomical
  Observatory of Padua
  \url{http://stev.oapd.inaf.it/cgi-bin/cmd\_2.3}} to model the mass
$M(g-r,\age,\feh)$ as a function of color $g-r$, age \age, and
metallicity \feh. The isochrones all assume $\afe = 0$. A lognormal
\citet{Chabrier01a} mass function $\phi(M)$ is used to weight the
contribution of various masses to the stellar population's mass (\ie,
the mass function provides the measure in mass space). Assuming
different forms for the IMF systematically changes the stellar
surface-mass densities by a few percent independent of metallicity.

The total stellar mass is related to the number of G-type dwarfs
approximately as
\begin{equation}
\mathrm{stellar\ mass\ in\ a\ \feh-\afe\ bin} = N(R_0;\feh,\afe)\, \langle M_\mathrm{G} \rangle(\feh) \, \omega^{-1}(\feh) \,,
\end{equation}
where $\langle M_\mathrm{G} \rangle$(\feh) is the average mass of a
G-type dwarf in the $0.45 \leq g-r \leq 0.58$ color
range\footnote{Here and in \eqnname s~(\ref{eq:mgfeh}) and
  (\ref{eq:avgmass}) we use a slightly wider color range than that of
  the \sdss/\segue\ G-dwarf sample, as the Padova isochrones do not
  have very high resolution in $g-r$ color.} and $\omega$(\feh) is the
ratio of a stellar population's mass in G-type dwarfs to the total
mass in the population. We calculate both $\langle M_\mathrm{G}
\rangle$(\feh) and $\omega$(\feh) using the stellar isochrones and
assuming a lognormal \citet{Chabrier01a} initial mass function.

We calculate the mass in G-type dwarfs of a stellar population as
\begin{equation}\label{eq:mgfeh}
M_{\mathrm{G,total}}(\feh) = \int_{0.45}^{0.58} \dd (g-r) \int_{\age_\mathrm{min}}^{\age_{\mathrm{max}}} \dd \age \int \dd M \,\phi(M)\,M(g-r,\age,\feh)\,,
\end{equation}
where we thus use a flat prior in age and we only use the dwarf part
of the isochrone. The total stellar mass of a stellar population is
given by a similar expression, but without the color restriction. In
what follows we use ($\age_{\mathrm{min}},\age_{\mathrm{max}}$) =
(0.5,10) Gyr for all abundance bins. As \afe\ is a relative age
indicator (see discussion in \bovyetal) with populations with $\afe
\gtrsim 0.25$ probably $\gtrsim 7$ Gyr old an \afe-dependent age
distribution would be more appropriate, although the exact form it
should take is hard to establish. We also used
($\age_\mathrm{min},\age_\mathrm{max}$) = (1,8) Gyr for $\afe < 0.25$
and ($\age_\mathrm{min},\age_{\mathrm{max}}$) = (7,10) Gyr for $\afe >
025$ and found no significant difference in the inferred stellar
masses for all of the abundance bins. $\omega(\feh)$ is then the ratio
of the stellar mass in G-type dwarfs to the total stellar mass. The
ratio $\omega(\feh)$ is approximately given by
\begin{equation}
\omega(\feh) \approx 0.0425 + 0.0198\,\feh\,+0.0057\,\feh^2\,,
\end{equation}
between -1.5 $< \feh < 0.3$.

We calculate the average mass of a G-type dwarf in our color range using a
similar procedure. We calculate the average mass by marginalizing over
the same age distribution, but without weighting by the mass function
as the color dependence of the mass function is very limited over this
narrow color range and the distribution of the \segue\ G-dwarf sample
is uniform in color $g-r$. The average mass is thus calculated as
\begin{equation}\label{eq:avgmass}
\langle M_\mathrm{G} \rangle (\feh) = \int_{0.45}^{0.58} \dd (g-r) \int_{\age_{\mathrm{min}}}^{\age_\mathrm{max}} \dd \age \, M(g-r,\age,\feh)\,;
\end{equation}
this is approximately given by
\begin{equation}
\langle M_\mathrm{G}/ M_\odot \rangle (\feh) 
 \approx 0.956 + 0.205\,\feh\ + 0.051\,\feh^2\,,
\end{equation}
again between -1.5 $< \feh < 0.3$.

The average mass $\langle M_\mathrm{G} \rangle (\feh)$ of a G-type dwarf as
a function of \feh\ only changes by about 20\,percent when going from
metal-poor to metal-rich stars, with more metal-poor stars having
smaller masses. The fractional contribution to the mass budget of G-type
dwarfs $\omega$(\feh) decreases from 5\,percent for metal-rich stars
to approximately 3\,percent for metal-poor stars. Thus, these
mass-correction factors are not the main drivers for the structure in
\figurename~\ref{fig:mass_hz_afe}.

\end{document}